\documentclass[aps,prl,twocolumn,superscriptaddress,showpacs,showkeys]{revtex4-1}
\usepackage{color}
\usepackage{amssymb}
\usepackage{amsmath}
\usepackage{bm}
\usepackage[normalem]{ulem}
\usepackage{graphicx}
\usepackage{epstopdf}
\usepackage{newtxtext,newtxmath}
\usepackage{xfrac}
\usepackage{amsfonts}

\newcommand{\cR}{{\mathcal{R}}}
\newcommand{\cC}{{\mathcal{C}}}
\newcommand{\tcR}{{\tilde{\mathcal{R}}}}
\newcommand{\tcC}{{\tilde{\mathcal{C}}}}
\newcommand{\tildet}{{\tilde{t}}}

\newcommand{\rmE}{{\rm E}} %% reaction set

\newcommand{\tmax}{{t_{\rm max}}}

\begin{document}
\setcounter{page}{1}
\title{Heterogeneous Popularity of Metabolic Reactions from Evolution}
\author{Mi Jin  \surname{Lee}}
\affiliation{Department of Applied Physics, Hanyang University, Ansan 15588, Korea}
\author{Deok-Sun \surname{Lee}}
\affiliation{School of Computational Sciences and Center for AI and Natural Sciences, Korea Institute for Advanced Study, Seoul 02455, Korea}
\email{deoksunlee@kias.re.kr}

\begin{abstract}
The composition of cellular metabolism is different across species. Empirical data reveal that bacterial species contain similar numbers of metabolic reactions but that the cross-species popularity of reactions is so heterogenous that some reactions are found in all the species while others are in just few species, characterized by a power-law distribution with the exponent one. Introducing an evolutionary model concretizing the stochastic recruitment of chemical reactions into the metabolism of different species at different times and their inheritance to descendants, we demonstrate that the exponential growth of the number of species containing a reaction and the saturated recruitment rate of brand-new reactions lead to the empirically identified power-law popularity distribution. Furthermore, the structural characteristics of metabolic networks and the species' phylogeny  in our simulations agree well with empirical observations.
\end{abstract}

%\date{\today}
\maketitle 

The orchestration of biochemical reactions to generate and consume matter and energy in cellular metabolism is essential for living organisms~\cite{michal1999biochemical,Jeong:2000wc}. Recently, thousands of species have had their genomes sequenced and annotated~\cite{10.1093/nar/gks1195}, enabling their reactions and biosynthetic and degradation pathways to be inferred computationally and databased~\cite{10.1093/nar/gkaa970,10.1093/bib/bbx085}.  The comparative and statistical analyses of the metabolic networks of such a larger number of different species disclose the current landscape of the cellular metabolism, enabling e.g., the phylogenetic analysis of metabolic pathway organizations~\cite{Mazurie:2008tb,Borenstein14482} and the analysis of different frequencies of individual reactions participating in the metabolism of species~\cite{Liu:2007un,Bernhardsson:2011ai,Kim:2015aa}.

How many species contain a given reaction in their metabolism, which we call popularity, represents how universally it is demanded. Some reactions execute crucial functions for most species and thus should be very popular, but others may be so only for a few species in special habitats. Therefore a difference in reactions' popularity may not be strange or surprising. Yet, as noted in ~\cite{Kim:2015aa} and will be investigated in detail here, the distribution of the reaction popularity exhibits a remarkable characteristics: it follows a power-law distribution with the exponent close to 1. This suggests that the reaction popularity is more broadly distributed than expected by chance and that, more importantly, it may be determined in a principled way---for instance,  intrinsically by its biochemical importance for life on earth or extrinsically built up over time under randomness. A plausible model that can reproduce this empirical finding can advance our understanding of the organizational principles of the cellular metabolism.

Here we show  that such heterogeneous popularity can be understood by studying the co-evolution of metabolism of individual species and the species tree. Previous studies on the metabolism evolution have considered a single abstracted metabolic network with new reactions and their catalytic enzymes added by some plausible mechanisms~\cite{Horowitz153,Ycas:1974dp,doi:10.1146/annurev.mi.30.100176.002205,Light:2004sp,Wagner:2009ul}.  
We consider a phylogeny of species and explore the possibility  that a reaction is dominantly found in the descendants of the ancestor species that first recruited it, and thus different first-recruitment times of reactions will result in different popularity in the contemporary species.   To be specific, we consider a growing species network where every node (species) contains a growing bipartite network of reactions and compounds, representing its metabolism, and such nodes may give birth to new nodes. In this model motivated by the recent study on the evolution of  ecological networks~\cite{maslov}, we reveal the core mechanism of diversifying the reaction popularity during metabolism evolution. We use the BioCyc database~\cite{10.1093/bib/bbx085} to predetermine the model parameters as much as possible. Simulating the model, we first obtain the popularity distribution that agrees excellently with the empirical one. Figuring out the time-dependent behaviors of key variables, we formulate the popularity distribution depending on the growing numbers of species and reactions. Moreover, this model not only explains reactions' heterogeneous popularity but also reproduces well the distribution of the distance between species, measured by the similarity of the set of reactions they have, and the degree distribution of individual metabolic networks, suggesting that it can play the role of a basic model for metabolism evolution. 

%%%%% Figure 1 %%%%%%%%%%%
\begin{figure}
\includegraphics[width=\columnwidth]{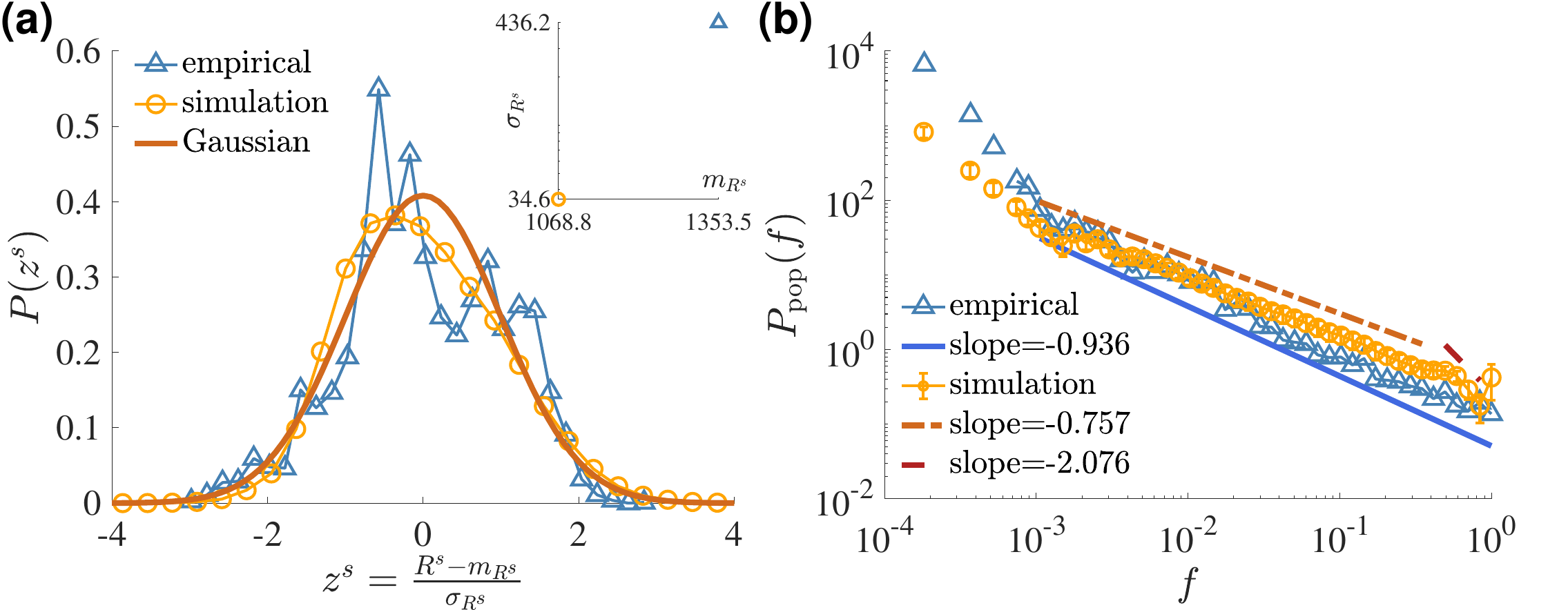}
\caption{Statistics of the species-reaction association in empirical data and the network evolution model with $\mu=0.02$.
(a) Standardized distributions of the number of reactions $R^s$ in a  species $s$.  The standardized variable $z^s = {R^s -m_{R^s} \over \sigma_{R^s}}$ is used with the mean $m_{R^s}$ and standard deviation $\sigma_{R^s}$ from empirical data (triangle) and from the model (circle). The solid line shows the Gaussian distribution ${1\over \sqrt{2\pi}}\exp\left(-{z^2\over 2}\right)$. (b) Distributions of the reaction popularity. The solid line fits the empirical data for $f>10^{-3}$, and the dashed-dotted and dashed lines fit the simulation results for $10^{-3}<f<0.4$ and $f>0.4$, respectively.}
\label{fig:reproduce}
\end{figure}
%%%%%%%%%%%%%%%%%%%%%%%%%%%%%

{\it Empirical results and a toy model.---}From the version 19.1 of the BioCyc database~\cite{10.1093/bib/bbx085}, we obtain the cellular metabolism of $5470$ bacterial species including a total of $R=11058$ chemical reactions. The number of reactions contained in a species is 
narrowly  distributed  following the Gaussian distribution with the mean $1354$ and standard deviation $436$ [Fig.~\ref{fig:reproduce}(a)]. This implies that most species adopt a similar size of metabolism although different environments may impose different constraints and demands that can be fulfilled by different pathways and reactions. Next, we define the popularity $f_r$ of a reaction $r$ as the fraction of species containing the reaction $r$~\cite{Bernhardsson:2011ai,Kim:2015aa}  and find that remarkably, its distribution is a power law with the exponent $1$ [Fig.~\ref{fig:reproduce}(b)]
\begin{equation}
P_{\rm pop} (f)\equiv{1\over R} \sum_{r} \delta(f_r-f) \sim f^{-1},
\label{eq:Pfemp}
\end{equation}
noting the higher abundance of popular reactions than expected under other distributions like an exponential one. We investigate how such strong heterogeneity and clear-cut power-law distribution emerge.

The depletion of resources and variation of environments can impose an evolutionary pressure facilitating the appearance of new enzymes catalyzing new reactions~\cite{Horowitz153,Ycas:1974dp,doi:10.1146/annurev.mi.30.100176.002205}. Those new reactions will be utilized by the species that first recruits them and by their descendants. These evolutionary processes can bring a power-law popularity distribution, as shown by the following toy model. 

Each species has a set of reactions for its metabolism. From time $t$ to $t+1$, every species $s$ gives birth to a daughter species $s^\prime$, which inherits all the reactions of $s$ and additionally recruits a new reaction $r_1$.  Simultaneously $s$ expands its metabolism by recruiting a new reaction $r_2$.  Therefore the  number of species increases with time $t$ as  $S(t)=2^t$ and the  number of distinct reactions $R(t)$ as $R(t+1)-R(t) = 2 S(t)$, giving $R(t) = 2^{t+1}-1$. 

Suppose that a reaction $r$ is recruited  by a species $s_r$ at time $\tau_r$ for the first time. It is found exclusively in $s_r$ and its descendants in this toy model. Therefore, $S_r(t) =  2^{t-\tau_r}$ species contain the reaction $r$ at time $t$, allowing us to compute its popularity   $f_r(t) =  {S_r(t) \over S(t)}=2^{-\tau_r}$. Notice that $f_r$ decays exponentially with the {\it first-recruitment time} $\tau_r$ but independent of the current time $t$. Among the $R(t)=2^{t+1}-1$ reactions present at time $t$, the fraction of reactions first recruited at $\tau \, (\leq t)$ is given by $P_{\rm rec}(\tau;t) \equiv {1\over R(t)} \sum_r \delta(\tau_r - \tau)= {R(\tau)-R(\tau-1)\over R(t)} = {2^{\tau} \over 2^{t+1}-1}$. Therefore, the reaction-popularity distribution $P_{\rm pop}(f;t)$ at time $t$ can be obtained from $P_{\rm rec}(\tau;t)$ and $f_r(t)$ as
\begin{equation} 
P_{\rm pop} (f;t) =\left|{df_r(t) \over d\tau_r}\right|^{-1}P_{\rm rec} (\tau_r;t)\bigg|_{f_r(t)=f}.
\label{eq:Pformula}
\end{equation}
For our toy model, ${df_r\over d\tau_r}|_{f_r=f} =-f \log(2)$ and $P_{\rm rec}(\tau_r;t)|_{f_r=f}= {f^{-1}\over 2^{t+1}-1}$, yielding $P_{\rm pop}(f)\sim f^{-2}$. Copying all reactions to the daughter species and recruiting a new reaction by every species at every step bring this power law. Yet the exponent $2$ is different from the empirical value $1$ [Eq.~(\ref{eq:Pfemp})], raising the need to improve the toy model. 
We use $t$ to denote the simulation time and $\tau$ to denote the first-recruitment time of a reaction. We will drop $t$ in $P_{\rm pop}(f;t)$ and $P_{\rm rec}(\tau;t)$  for simplicity.  

%%%%% Figure 2: Model %%%%%%%%%%%
\begin{figure}
\includegraphics[width=0.95\columnwidth]{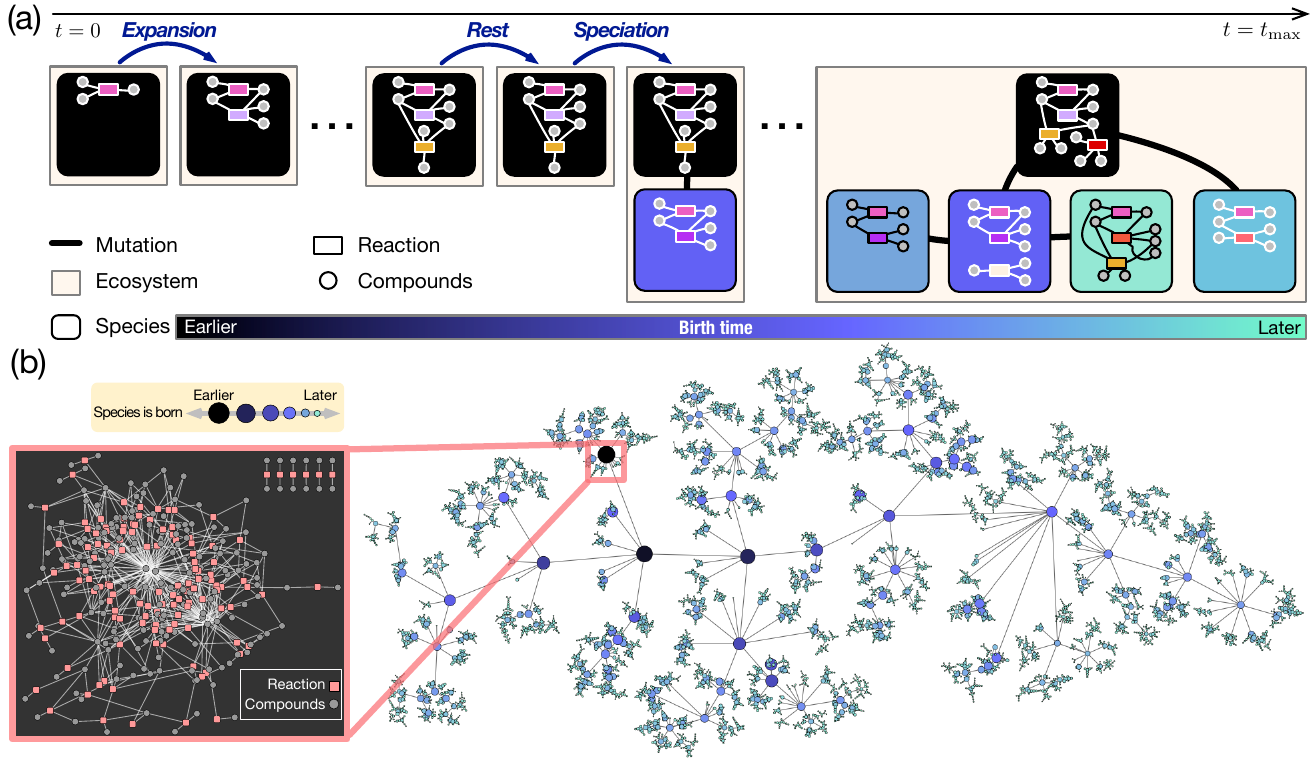}
\caption{Network evolution model.  (a) A bipartite network of reactions (rectangles) and compounds (circles) represents each species. At every time step, each species may do nothing (rest) or evolve by either gaining a new reaction (expansion) or giving birth to a daughter species inheriting active components formed by a new reaction replacing an old one (speciation). 
(b) The species tree from a simulation with $\mu=1$  is shown, where nodes represent 5660 species and links represent the parent-daughter relationship. Node size and color vary with the birth time of the corresponding species. The metabolic network of the oldest species is shown in the left box.  }
\label{fig:model}
\end{figure}
%%%%%%%%%%%%%%%%%%%%%%%%%%%%%

{\it Network evolution model.---}Differently from the toy model,  the pool of reactions to recruit is finite, and the same reaction can be recruited by more than one species. Only the reactions that could be activated, i.e., maintain a nonzero flux, are recruited by a species~\cite{Handorf2005,maslov,SM}. Therefore, we need to consider the connection of the recruited reactions  in each species.  The birth rate of a new species by mutation, replacing a current reaction by a similar but more competent reaction, may be different from the expansion rate of individual metabolism.

To incorporate these aspects, we consider a growing species tree with each species possessing a growing bipartite network of reactions and compounds~\cite{metabolic_network} as shown in Fig.~\ref{fig:model}.  Initially ($t=0$), a single species is born, possessing a bipartite network of a single {\it stand-alone} reaction and its compounds. A stand-alone reaction is one that takes in externally available compounds and therefore can be activated independently without requiring the activation of other reactions, and we have 440 such stand-alone reactions~\cite{SM}. From $t$ to $t+1$,  for each species $s$, a potential new reaction $r_{\rm new}$ is selected from the reaction pool, which takes in the compounds already included in $s$  or externally available compounds. Then we investigate whether there is a reaction $r_{\rm sim}$ in $s$ similar to $r_{\rm new}$, sharing the same set of substrates or products~\cite{SM}. If there is no such similar reaction, then  $s$ recruits $r_{\rm new}$  (expansion).  Otherwise, the species $s$ either gives birth to a new species or does nothing as follows. If the competence $\phi_{r_{\rm new}}$ of the potential new reaction, assigned randomly to each reaction, is higher than that of the similar reaction $\phi_{r_{\rm sim}}$, then a new species $s_{\rm new}$ is born with probability $\mu$, inheriting the active connected components of $s$ formed after replacing  $r_{\rm sim}$  by $r_{\rm new}$ in $s$ (speciation). Otherwise, nothing happens (rest).  These procedures are sketched in Fig.~\ref{fig:model}(a). Here a connected component  is considered active if it  contains at least one stand-alone reaction so that it can maintain a nonzero flux.

\begin{figure}
\includegraphics[width=\columnwidth]{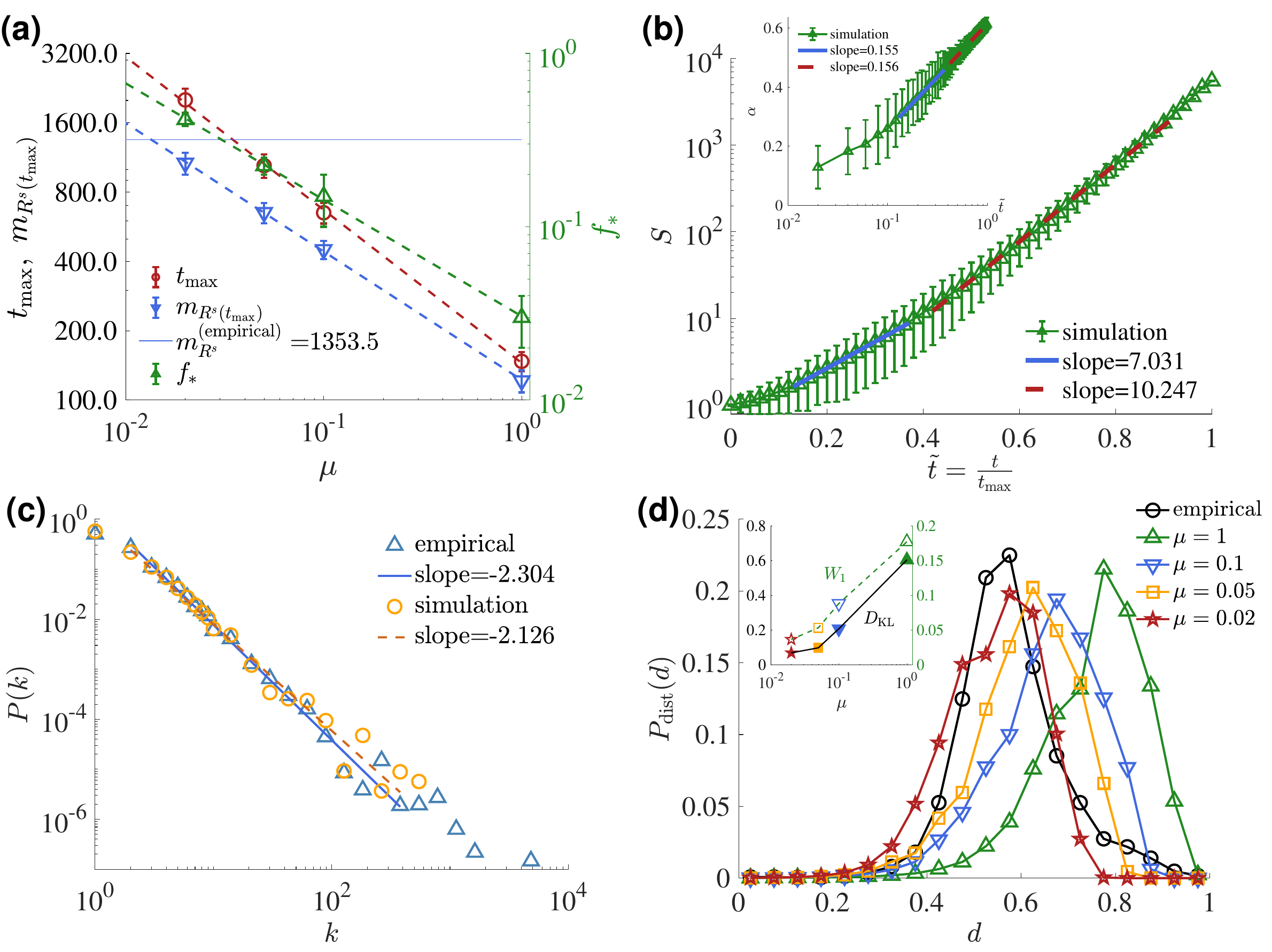}
\caption{Growth of the species tree.
(a) Simulation time $t_{\rm max}$ to generate $S\geq 5470$ species, the mean number of  reactions in a species $m_{R^s(t_{\rm max})}$, and the crossover scale of popularity $f_*$ versus the speciation rate $\mu$. Dashed lines fit the data, respectively. The empirical value  $m_{R^s}^{\rm (empirical)}=1353.5$ is expected at $\mu^{\rm (empirical)}\simeq 0.0134$. 
(b) The number of species $S$ versus the normalized time $\tildet = {t\over \tmax}$ for $\mu=0.02$.  Inset: the probability $\alpha$ that a potential reaction finds a similar reaction in a species versus $\tilde{t}$.  The solid and dashed lines fit the data for $\tilde{t}\leq \tildet_*\simeq 0.4$ and $\tildet>\tildet_*$, respectively.  
(c) Distributions of the degree of the compounds from the empirical data and simulations with $\mu=0.02$.
(d) Distributions of the metabolic distance of two species from the empirical data and simulations with different $\mu$'s. Inset: KL divergence and 1-Wasserstein distance versus $\mu$.
 }
\label{fig:temporal}
\end{figure}

We simulated this model until $\tmax$ when the number of species exceeds the empirical value, i.e., $S(\tmax)\geq 5470$. Figure~\ref{fig:model}(b) shows the obtained tree of metabolic networks. The parameter $\mu$ controls the rate of speciation. Furthermore, with increasing $\mu$, the simulation time $t_{\rm max}$  and the mean number of reactions per species decrease [Fig.~\ref{fig:temporal}(a)]. The empirical value $1354$ is expected  at $\mu=\mu^{\rm (empirical)}\simeq 0.013$, while the computing time and resource constraints limit our simulation to $\mu\geq 0.02$.

The popularity distribution $P(f)$ from the model takes a power law with the exponent close to one in a wide range of $f$ in agreement with the empirical result [Fig.~\ref{fig:reproduce}(b)].  Interestingly, a crossover to faster decay with the exponent close to $2$ is observed for large $f$ as 
\begin{equation}
P_{\rm pop}(f) \sim 
\begin{cases}
f^{-\eta_1} & \ {\rm for} \ f<f_*,\\
f^{-\eta_2} & \ {\rm for} \ f>f_*
\end{cases}
\label{eq:Pfcrossover}
\end{equation}
with $(\eta_1,\eta_2)=(0.757, 2.08)$ and the crossover scale  $f_*\simeq 0.42$ for $\mu=0.02$. The large-$f$ regime, characterized by the exponent $2$ of the toy model, is narrow and shrinks  as $\mu$ decreases; $f_*$ is $0.56$ at $\mu^{\rm (empirical)}$.  The power-law exponents vary rarely with $\mu$~\cite{SM}.  Below we investigate the time dependence of the key quantities and various results from this network model to understand how it reproduces the empirical findings better than the toy model and why it displays the crossover behavior.

%%%%% Figure 3:  %%%%%%%%%%%
%%%%%%%%%%%%%%%%%%%%%%%%%%%%%

{\it Structure of species tree and metabolic networks.---}	The total number of species grows as $\langle S(t+1)\rangle - \langle S(t)\rangle \simeq {1\over 2} \mu \alpha(t) \langle S(t)\rangle$, where $\langle \cdots\rangle$ is the ensemble average,  $\alpha(t)$ is the probability that a potential new reaction $r_{\rm new}$  has a similar reaction $r_{\rm sim}$ in a species, and $1/2$ is the probability of $\phi_{r_{\rm new}}>\phi_{r_{\rm sim}}$.  $\alpha(t)$ grows very weakly (logarithmically), and therefore $\langle S\rangle \sim \exp({1\over 2}\mu \bar{\alpha}t)$. We estimate $\bar{\alpha}=0.40$ and $0.55$  for $\tildet<\tildet_*$ and $\tildet>\tildet_*$, respectively, with  the bar meaning time average, the normalized time $\tildet\equiv {t\over \tmax}$, and the normalized crossover time $\tildet_*=0.40$ (for $\mu=0.02$) distinguishing  the early- and late-time regimes showing different behaviors of $S(t)$ and related to  $f_*$~\cite{SM}.

The structure of the metabolic networks of individual species agrees with the empirical data. In Fig.~\ref{fig:temporal}(c), the degree distribution of compounds from simulations is shown to coincide with the empirical one, both taking a power law with the exponent close to $2$. To compare the simulated phylogeny with the real-world one, we introduce the metabolic distance $d_{s_1,s_2}$ between two species defined as $1-$ (the relative size of the intersection with respect to the union of their sets of reactions)~\cite{SM} and obtain its distribution. When $\mu=0.02$, the obtained distance distribution agrees well with the empirical one [Fig.~\ref{fig:temporal}(d)]. A reasonable agreement is expected also at  $\mu^{\rm (empirical)}\simeq 0.013$ given the reduced decrease of the Kullback-Leibler (KL) divergence and the 1-Wasserstein distance~\cite{Villani2009,SM} with decreasing $\mu$ around $\mu=0.02$.
%, the model distribution  the empirical one as $\mu$ decreases, as quantified by  so that they agree well at $\mu=0.02$ and are likely to do so also at $\mu^{\rm (empirical)}$, implying the agreement of their phylogeny} 

%%%%% Figure 4: two phases %%%%%%%%%%%
\begin{figure}
\includegraphics[width=\columnwidth]{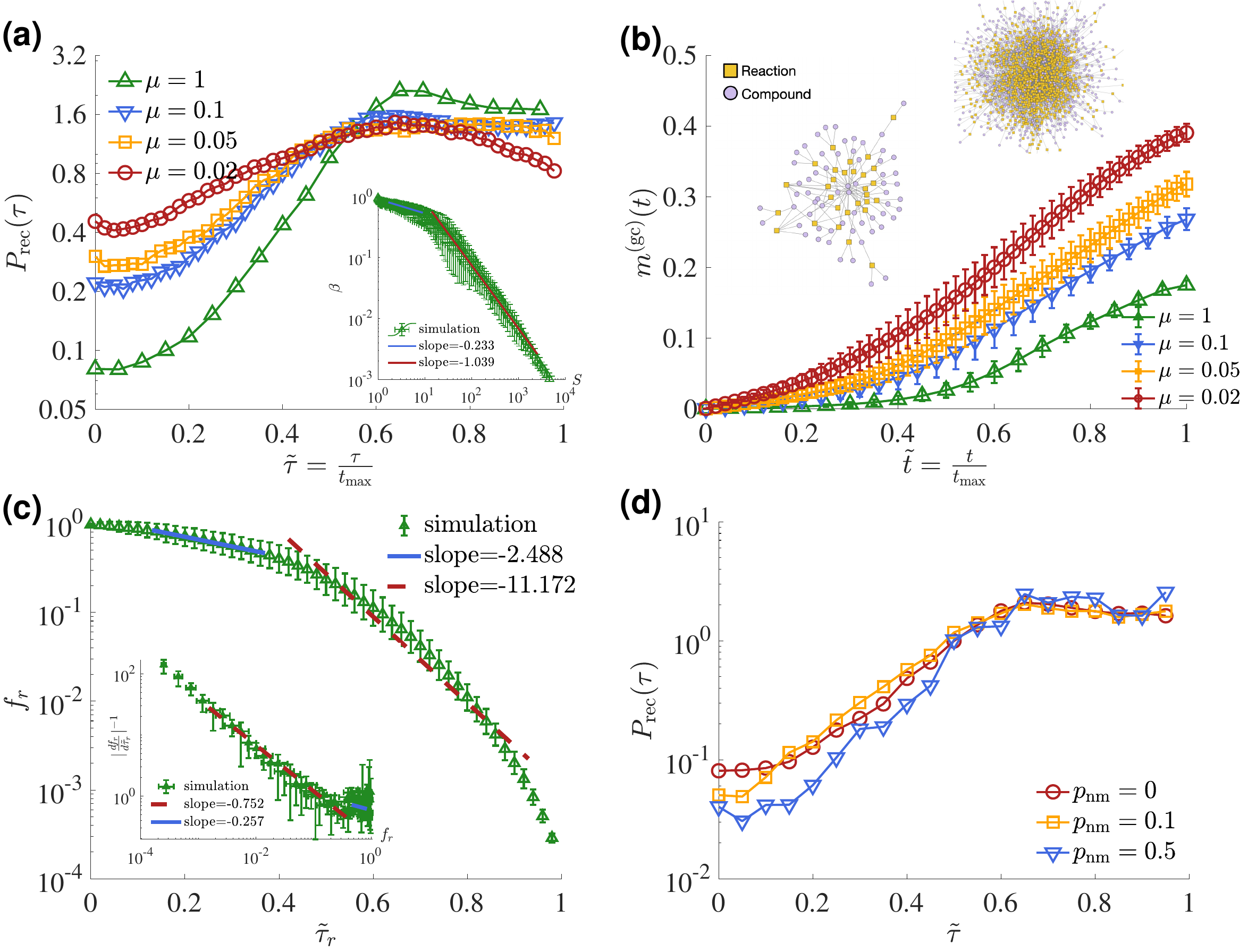}
\caption{First-recruitment and popularity of reactions in the network evolution model.  (a)  Distribution of the normalized first-recruitment time $\tilde{\tau}_r = {\tau_r \over \tmax}$ of a reaction $r$ with different $\mu$'s. 
Inset: plot of the probability $\beta(\tau)$ that a reaction recruited at time $\tau$ is brand-new versus the total number of species $S(\tau)$ with $\mu=0.02$. 
(b) Time evolution of the fraction of distinct recruited reactions in the largest component $m^{\rm (gc)}$ in the universal reaction-compound network. Also shown are the largest components at $\tildet \simeq 0.18$ and $\tildet\simeq 0.69$, respectively. 
(c) The popularity $f_r$ versus $\tilde{\tau}_r$ with $\mu=0.02$. Inset: plot of $|{df_r\over d\tau_r}|^{-1}$ versus $f_r$. Dashed and solid lines fit the data for $f_r<f_*$ and $f_r>f_*$, respectively. 
(d) First-recruitment time distributions for $\mu=1$ and different probabilities of nonmetabolic speciation $p_{\rm nm}$. 
}
\label{fig:twophases}
\end{figure}
%%%%%%%%%%%%%%%%%%%%%%%%%%%%%

{\it First recruitment and popularity of reactions.---}A reaction just recruited by a species may not be {\it brand-new} but has already been recruited by some other species. The union of the sets of the reactions recruited by all the present species,  the size of which we denote by $R(t)$, expands only when a species recruits a brand-new reaction. Let us consider $\beta(\tau)$ the probability that the reaction recruited at time $\tau$ by a species is brand-new. Among all the reactions present at time $t_{\rm max}$,  the fraction of the reactions first recruited at $\tau$, or  the distribution of the first-recruitment time of a reaction, is given by $P_{\rm rec}(\tau)\simeq  \left[1 - \alpha(\tau) + {1\over 2} \mu \alpha (\tau) \right] \beta(\tau) \langle S(\tau)\rangle/\langle R(\tmax)\rangle$, where $1-\alpha+(1/2)\mu\alpha$ is the probability of expansion or speciation. If $\beta(\tau)$ varies weakly, $P_{\rm rec}(\tau)$ will grow exponentially with $\tau$. 

Simulations show that $P_{\rm rec}(\tau)$ first grows exponentially with $\tau$ and then saturates or decreases weakly with $\tau$ as shown in Fig.~\ref{fig:twophases}(a). With $\mu=0.02$, we find
\begin{equation}
P_{\rm rec}(\tau) \sim 
\begin{cases}
\exp(2.86\, \tilde{\tau}) & \ {\rm  for} \  \tilde{\tau}\equiv{\tau\over \tmax}\lesssim \tildet_*= 0.40,\\
{\rm varying \ weakly} & \ {\rm for} \ \tilde{\tau}\gtrsim \tildet_*.
\end{cases}
\label{eq:Precnet}
\end{equation}
Consequently, $R(t) = R(\tmax)\sum_{\tau<t} P_{\rm rec}(\tau;t_{\rm max})$ grows  first exponentially and then linearly with time $t$, distinguishing the two time regimes. For $\tilde{\tau}>\tilde{t}_*$, despite the exponentially growing $S(\tau)$ and as much frequent expansion of individual metabolic networks, most  recruited reactions are not brand-new; the brand-new probability $\beta(\tau)$ scales inversely with $S(\tau)$ as
\begin{equation}
\beta(\tau)\sim 
\begin{cases}
S(\tau)^{-0.233} & {\rm for} \ \tilde{\tau}\lesssim \tilde{t}_*\\
S(\tau)^{-1.04} & {\rm for} \ \tilde{\tau}\gtrsim \tilde{t}_*,
\end{cases}
\end{equation}
as shown in the inset of Fig.~\ref{fig:twophases}(a). Therefore, $\beta(\tau) S(\tau)$ is almost independent of $\tau$, allowing us to understand the saturation of $P_{\rm rec}(\tau)$ in the late-time regime. Underlying these phenomena is that the ever-recruited reactions and their compounds form the giant (percolating) component in the universal reaction-compound network. The portion of reactions in the  component $m^{(gc)}$ is of order 1 in the late-time regime [Fig.~\ref{fig:twophases}(b)], and thus a potential new reaction for each species is increasingly likely to overlap with the giant component, resulting in the decaying $\beta$.

A reaction $r$  recruited first by a species $s_r$ at time $\tau_r$ can be found at later times in the  descendants of $s_r$ inheriting $r$ and also in other species recruiting $r$ later than $s_r$.  The number $S_r^{\rm (0)}(t)$ of the descendants  of $s_r$ containing $r$ is a lower bound for $S_r(t)$ and  analyzed as follows.  Let $\omega(t)$ denote the probability that a reaction belongs to an active component after a  reaction is replaced by another. Then, $S_r^{(0)}(t)$ satisfies  $\langle S^{(0)}_r (t+1)\rangle - \langle S^{(0)}_r(t)\rangle = {1\over 2} \mu \alpha (t) \omega(t) \langle S^{(0)}_r(t)\rangle$,  leading to $\langle S^{(0)}_r(t)\rangle \sim \exp[{1\over 2} (t - \tau_r) \mu \bar{\alpha} \bar{\omega} ]$ and $f_r^{(0)}(\tmax) = {\langle S^{(0)}_r(\tmax)\rangle\over S(\tmax)}\propto \exp[- {1\over 2} \mu\bar{\alpha} \bar{\omega}  \tau_r]$. In a reasonable agreement with this prediction,  $f_r$ behaves approximately as [Fig.~\ref{fig:twophases}(c)] 
\begin{equation}
f_r  \sim  
\begin{cases}
\exp(-2.49\, \tilde{\tau_r}) & \ {\rm  for} \  \tilde{\tau_r}\equiv{\tau_r \over \tmax} \lesssim\tildet_*,\\
\exp(-11.2\, \tilde{\tau_r}) & \ {\rm for} \ \tilde{\tau_r}\gtrsim\tildet_*,
\end{cases}
\label{eq:fnet}
\end{equation}
where the faster decay  is  related to the larger value of $\bar{\alpha}$ and $\bar{\omega}$. 

These results and the formula in Eq.~(\ref{eq:Pformula}) enable us to see that the power-law decay of $P_{\rm pop}(f)$  is mainly attributed to the exponential decay of $f_r$ with $\tau_r$  in Eq.~(\ref{eq:fnet}) and its crossover originates in $P_{\rm rec}(\tau_r) \sim f_r^{-1}$ and $P_{\rm rec}(\tau_r)\sim {\rm const.}$ in the early- and late-time regimes, respectively. The results for the late-time regime apply to almost all reactions and species and therefore its agreement with the empirical data is remarkable. The region of $f$ displaying the faster decay $P_{\rm pop}(f)\sim f^{-2}$, absent in the empirical data, shrinks with decreasing $\mu$, and can be further diminished by e.g., allowing earlier-born species to recruit a larger number of reactions, which can make $P_{\rm rec}(\tau)$ vary weakly with $\tau$ in the whole time regime. More improvements toward reality can be pursued. 

{\it Discussion.---}We have studied the origin of the power-law distribution of the metabolic reaction popularity by investigating a network co-evolution model. The birth of a new species inheriting the metabolic network of its parent species and its expansion by recruiting reactions can generate such heterogeneity in the reaction popularity as observed empirically. We investigated the time dependence of the numbers of species and distinct reactions, and the popularity of individual reactions. Contrary to the toy model,  the reaction pool is finite, and the total number of distinct recruited reactions grows linearly with time in the late-time regime,  which brings the popularity distribution with the exponent one in agreement with the empirical distribution. In addition, the structure of individual species' metabolic networks and the phylogeny of the species from the model simulations agree well with the empirical data, demonstrating the potential of our model to be an elementary model for the metabolism evolution.

Varying the size of the reaction pool or the composition of the stand-alone reaction set does not change the main results. While a new species can be born by non-metabolic pressure,  the inclusion of a  nonmetabolic speciation with probability $p_{\rm nm}$ in our model, by which a new species is born with the same metabolic network as its parent, does  not change the results qualitatively [Fig.~\ref{fig:twophases}(d)].  Our model considers only the growth mechanism, but  to be more realistic,  the retirement of existing reactions and species extinction may be considered.  It will be interesting to study the statistics of the traits hitchhiking in the metabolism evolution in our model.

%%%% Acknowledgment
\begin{acknowledgments}
This work was supported by the National Research Foundation of Korea (NRF) grants funded by the Korean Government [Grants No. 2021R1C1C1007918 (M.J.L) and 2019R1A2C1003486 (D.-S.L)] and a KIAS Individual Grant (No. CG079901) at Korea Institute for Advanced Study (D.-S.L). This work is supported by the Center for Advanced Computation at Korea Institute for Advanced Study.
\end{acknowledgments}

%\bibliography{metabolic}

\onecolumngrid
\begin{center}
{\bf \large Supplemental Material}
\end{center}

\renewcommand{\thesection}{S\arabic{section}}
\renewcommand{\theequation}{S\arabic{equation}}
\renewcommand{\thefigure}{S\arabic{figure}}
\renewcommand{\thetable}{S\arabic{table}}
\setcounter{figure}{0}    
\setcounter{equation}{0}

\section{Empirical data and network evolution model}

From the version 19.1 of the BioCyc database~\cite{10.1093/bib/bbx085}, we obtain the matrix $G_c^r$ representing the stoichiometric coefficients of compound $c$ in reaction $r$ as well as the species-reaction association matrix $A^s_r=1$ or $0$,  representing whether a species $s$ contains a metabolic reaction $r$  or not,  for $S=5470$ bacterial species and $R=11058$  reactions, and $C=7620$ compounds. The number of reactions $R^s\equiv \sum_r A^s_r$ contained in (the metabolism of) a single species $s$ is narrowly distributed  following the Gaussian distribution with the mean $m_{R^s}\equiv \sum_s R^s/S\simeq 1353.5$ and standard deviation $\sigma_{R^s} \equiv [\sum_s (R^s - m_{R^s})^2/S]^{1/2}\simeq  436.2$ [Fig. 1(a)].

In the network evolution model, we update the set of the reactions that can be potentially recruited,  $\tcR^s(t)\equiv \{r\in \tcR -  \cR^s(t)| \cC_{r-}\subset (\cC^s(t) \cup \tcC^\rmE) \ {\rm or} \ \cC_{r+}\subset (\cC^s(t) \cup \tcC^\rmE) \}$, at every step $t$ for every species $s$, with  $\tcR$ a universal pool of  $R=11058$ reactions. $\cR^s(t) \ [\cC^s(t)]$ is the set of the reactions (compounds) contained in species $s$, $\cC_{r-(+)}$ denotes the set of the substrates (products) of reaction $r$, and $\tcC^\rmE$ is the set of $C^\rmE=138$ externally available compounds known for the flux-balance modeling of the metabolism of {\it E. coli}~\cite{Feist2007}. Every reaction is assumed to be reversible, and thus  the distinction between the set of substrates and products is arbitrary; The set $\cC_{r-}$ of substrates and $\cC_{r+}$ of products of reaction $r$ can be exchanged. We call two reactions $r_1$ and $r_2$ similar if they share the same set of substrates or of products, i.e., $\{\cC_{r_1-}, \cC_{r_1+}\} \cap \{\cC_{r_2-}, \cC_{r_2+}\}\neq \emptyset$.  Every reaction $r$ is assigned competence $\phi_r$ distributed uniformly.

Initially ($t=0$), a single species is born, possessing a bipartite network of a single reaction selected  from  the set of stand-alone reactions $\tcR^{\rm (sa)}\equiv \{r|\cC_{r-}\subset \tcC^\rmE \ {\rm or} \ \cC_{r+}\subset \tcC^\rmE \}=\tcR^s(0)$ and its compounds. From $t$ to $t+1$,  a potential new reaction $r_{\rm new}$ is selected from $\tcR^s(t)$ for each  $s$.  If there is no similar old reaction $r_{\rm sim}$ in $s$, then  $s$ recruits $r_{\rm new}$  (expansion).  Otherwise, the species $s$ either gives birth to a new species or does nothing as follows. If $\phi_{r_{\rm new}}> \phi_{r_{\rm sim}}$, then a new species $s_{\rm new}$ is born with probability $\mu$, inheriting the active connected components of $s$ formed after replacing  $r_{\rm sim}$  by $r_{\rm new}$ in $s$ (speciation). Otherwise, nothing happens (rest).  These procedures are sketched in Fig. 2(a). Here a connected component is considered active if it  contains at least one stand-alone reaction processing the externally available substrates or products, and thus can maintain a non-zero flux.

The motivation of defining the set of potential reactions $\tcR^s(t)$ as given in the main text is as follows. For a species $s$ having a set of recruited reactions $\cR^s(t) \equiv \{r|A^s_r(t)>0\}$ and of their compounds $\cC^s(t)\equiv \{c| A^s_c(t) \equiv  \theta(\sum_r A^s_r |G^r_c|)>0\}$, where $\theta(x)=1$ for $x>0$ and $0$ otherwise, one can expect that the compounds in the union of $\cC^s(t)$ and $\tcC^\rmE$ are available for $s$, as they are available already in $s$ or present externally.  Therefore every new reaction whose whole substrates or  products belong to $\cC^s(t) \cup \tcC^\rmE$ can be activated, i.e., have a non-zero flux in principle when added to $s$. This reasoning leads us to define  the potential reaction set as  $\tcR^s(t)\equiv \{r\in \tcR -  \cR^s(t)| \cC_{r-}\subset (\cC^s(t) \cup \tcC^\rmE) \ {\rm or} \ \cC_{r+}\subset (\cC^s(t) \cup \tcC^\rmE) \}$.

\section{Varying the speciation rate parameter $\mu$}
\label{sec:app_mus}

The parameter $\mu$, the only parameter of the network evolution model, controls the speed of speciation and affects the number of reactions per species and other properties as shown in Fig. 3(a). We take $\mu=1, 0.1, 0.05$, and $0.02$ within the limit of computation resource and time, and present the results obtained with $\mu=0.02$, which is the closest to the empirical values $\eta^{\rm (empirical)}\simeq 0.013$ estimated based on the dependence of the mean number of reactions per species. Here we present the results obtained with $\mu=1, 0.1$, and $\mu=0.02$ for comparison with the results in the main text.

%%%%% Figure S1rev:  %%%%%%%%%%%
\begin{figure}
\includegraphics[width=\columnwidth]{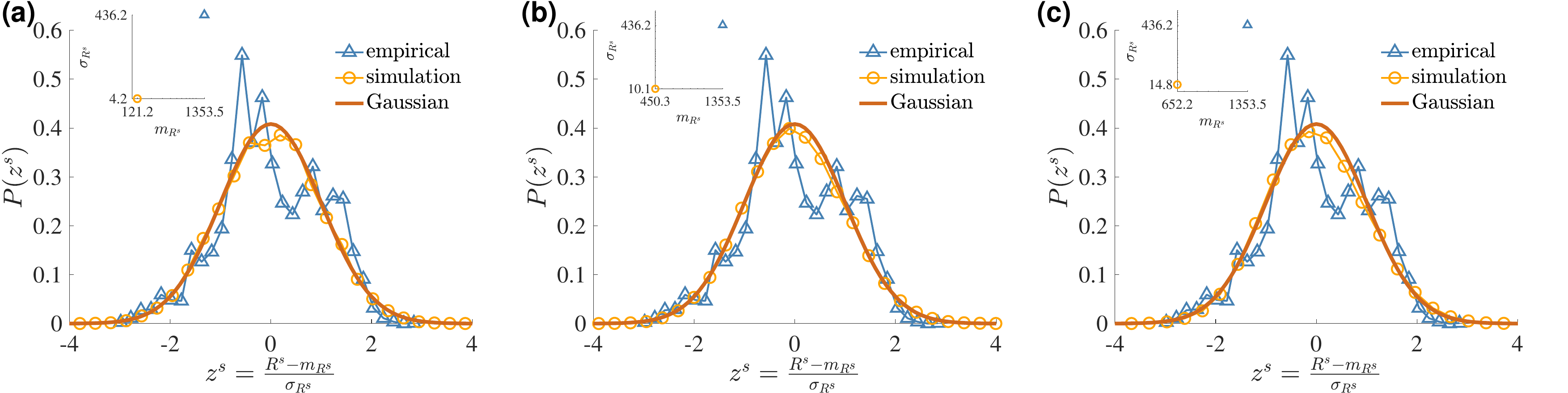}
\caption{ Standardized distributions of the number of reactions in a  species in the network evolution model for (a) $\mu=1$, (b) $\mu=0.1$, and (c) $\mu=0.05$. }
\label{fig:PRSmu}
\end{figure}
%%%%%%%%%%%%%%%%%%%%%%%%%%%%%

%%%%% Figure S2rev:  %%%%%%%%%%%
\begin{figure}
\includegraphics[width=\columnwidth]{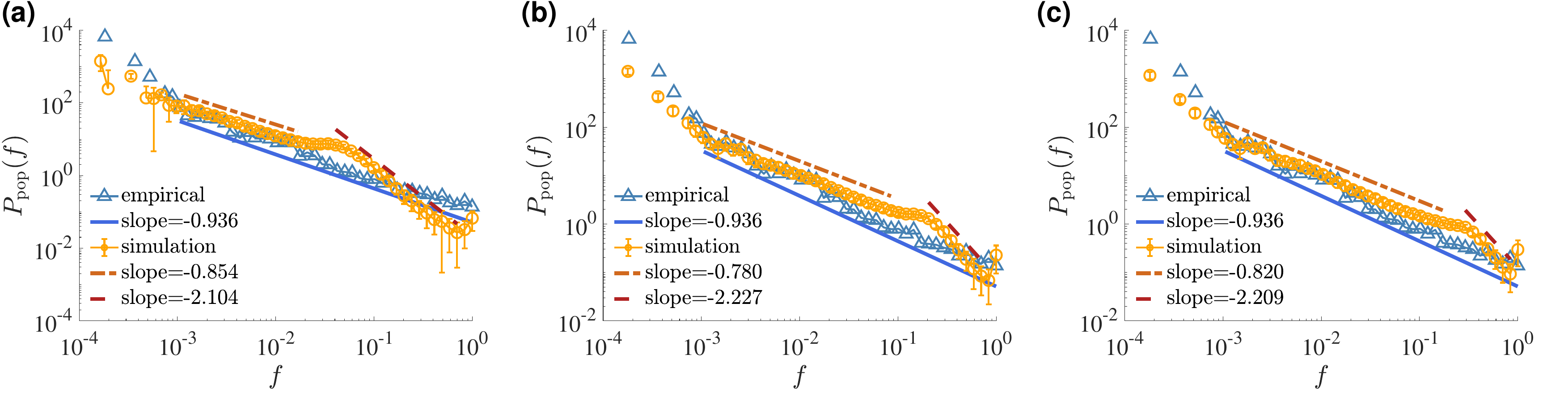}
\caption{ Distributions of the popularity $f$ of a reaction in the network evolution model for (a) $\mu=1$, (b) $\mu=0.1$, and (c) $\mu=0.05$.}
\label{fig:Ppopfmu}
\end{figure}
%%%%%%%%%%%%%%%%%%%%%%%%%%%%%

%%%%% Figure S3rev:  %%%%%%%%%%%
\begin{figure}
\includegraphics[width=\columnwidth]{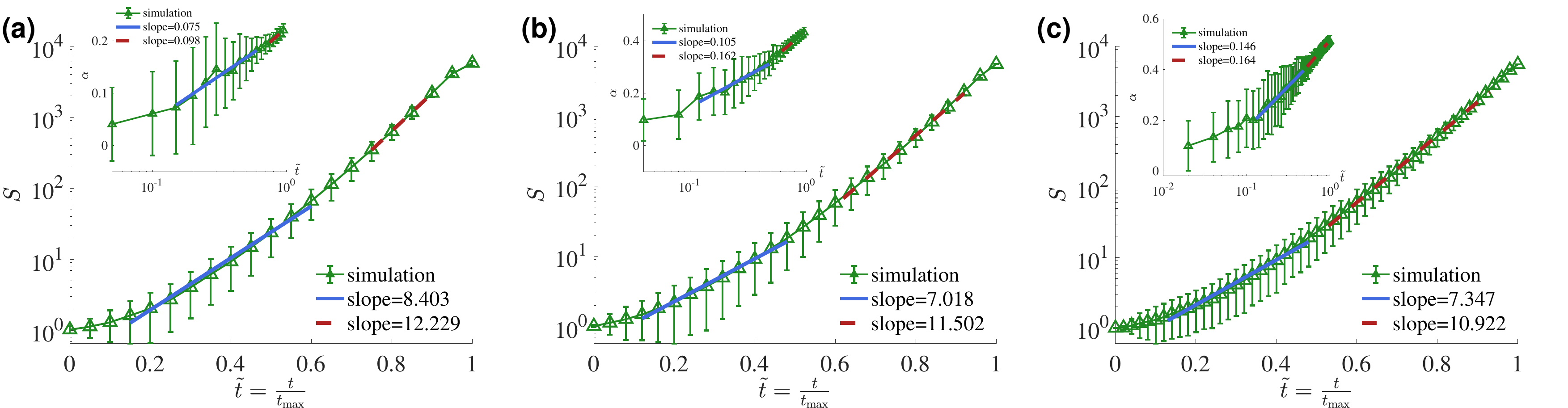}
\caption{ The number of species $S$ versus the normalized time $\tildet = {t\over \tmax}$ for (a) $\mu=1$, (b) $\mu=0.1$, and (c) $\mu=0.05$.  Inset: The probability $\alpha$ that a potential reaction finds a similar reaction in a considered species versus $\tilde{t}$.  The solid and dashed lines fit the data for $\tilde{t}\leq \tildet_*$ and $\tildet>\tildet_*$, respectively. }
\label{fig:Staumu}
\end{figure}
%%%%%%%%%%%%%%%%%%%%%%%%%%%%%

%%%%% Figure S4rev:  %%%%%%%%%%%
\begin{figure}
\includegraphics[width=\columnwidth]{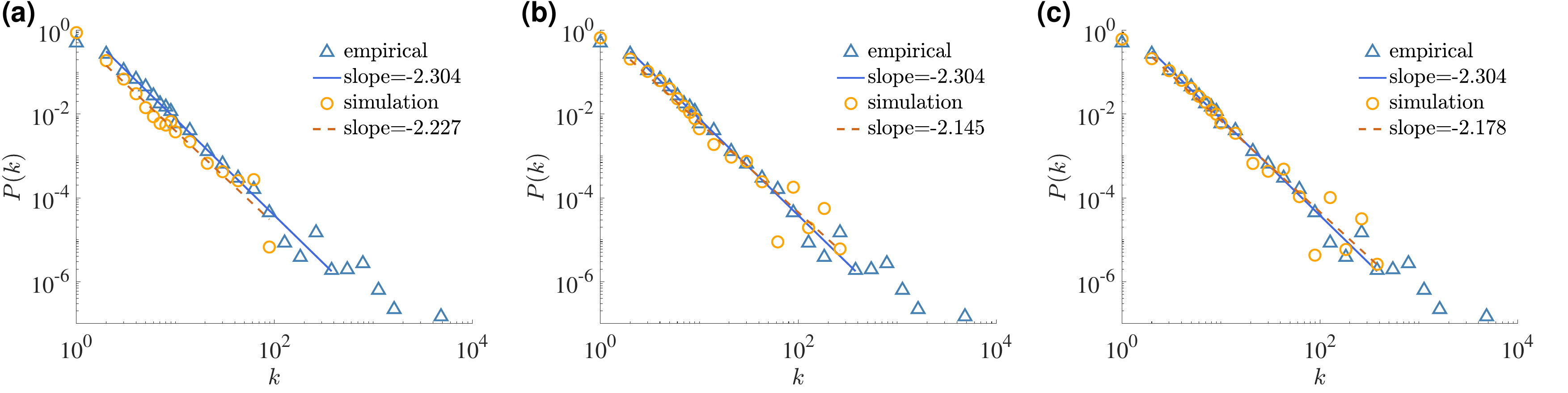}
\caption{Degree distributions from the empirical data and simulations with (a) $\mu=1$, (b) $\mu=0.1$, and (c) $\mu=0.05$.  }
\label{fig:Pkmu}
\end{figure}
%%%%%%%%%%%%%%%%%%%%%%%%%%%%%

%%%%% Figure S5rev:  %%%%%%%%%%%
\begin{figure}
\includegraphics[width=\columnwidth]{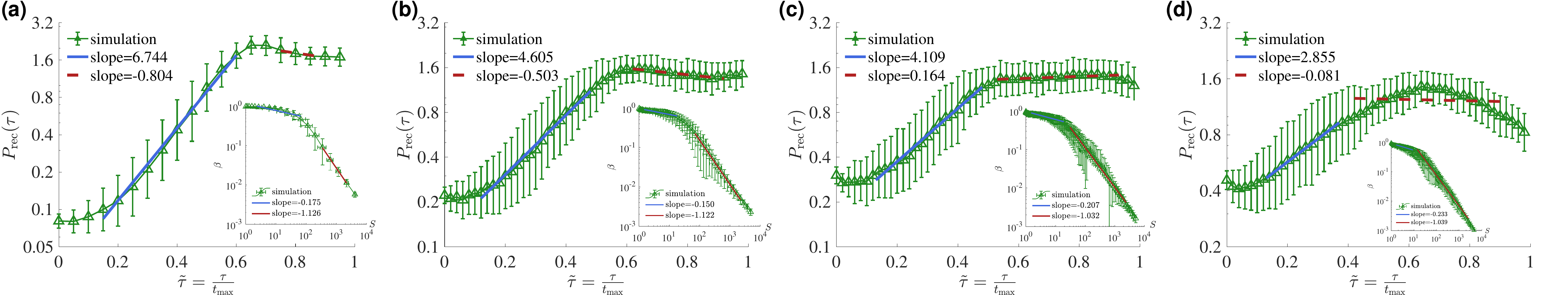}
\caption{Distribution of the normalized first-recruitment time $\tilde{\tau} = {\tau \over \tmax}$ of a reaction $r$ for (a)$\mu=1$, (b) $\mu=0.1$, (c)$\mu=0.05$, and (d)$\mu=0.02$. Solid and dashed lines fit the data for $\tilde{\tau}\leq \tildet_*$ and $\tilde{\tau}_r>\tildet_*$, respectively. Inset: Plot of the probability $\beta$ that a potential reaction  is brand-new versus the total number of species.  }
\label{fig:Precmu}
\end{figure}
%%%%%%%%%%%%%%%%%%%%%%%%%%%%%

%%%%% Figure S6rev:  %%%%%%%%%%%
\begin{figure}
\includegraphics[width=\columnwidth]{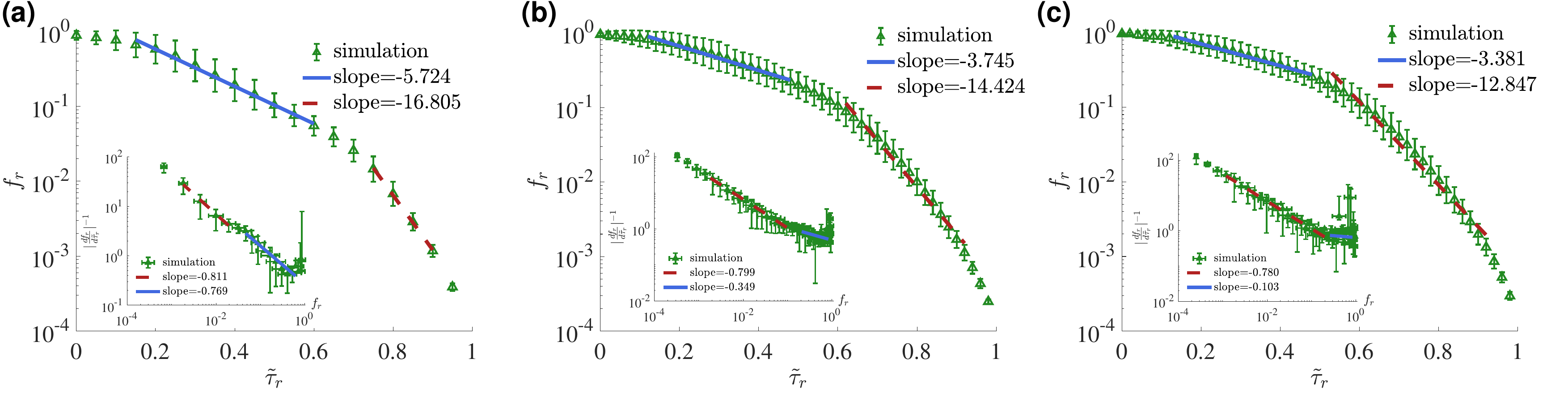}
\caption{ The popularity $f_r$ of a reaction $r$ versus its normalized first-recruitment time $\tilde{\tau}_r$. Inset: The inverse of its derivative versus popularity. Dashed and solid lines fit the data for $f_r<f_*$ and $f_r>f_*$, respectively. }
\label{fig:ftaumu}
\end{figure}
%%%%%%%%%%%%%%%%%%%%%%%%%%%%%

\begin{itemize}
\item
Regardless of the value of $\mu$, the standardized distribution of the number of reactions per species takes the Gaussian form as shown in Fig.~\ref{fig:PRSmu}. The mean and standard deviation decrease with increasing $\mu$ as shown in Fig. 3(a) in the main text. 
\item
The large-$f$ regime displaying a faster decay than the small-$f$ regime is wide for $\mu=1$, but shrinks as $\mu$ decreases [Fig.~\ref{fig:Ppopfmu}]. We identify the boundary $f_*$between the two regimes, the small- and large-$f$ regimes, by inspecting the simulation data, which is 
$f_*=0.03\pm 0.01, 0.15 \pm 0.005, 0.225\pm 0.025$, and $0.415\pm 0.035$ for $\mu  = 1, 0.1, 0.05$, and $0.02$, respectively. 
\item
The exponential growth of the number of species $S(t)$ with time $t$ is preserved across $\mu$ with the exponential growth rate slightly varying with $\mu$ [Fig.~\ref{fig:Staumu}]. The probability $\alpha$ that a potential reaction finds a similar reaction in a considered species grows logarithmically with time and the coefficient is similar between the early- and the late-time regime.  The normalized crossover time $\tilde{t}_* = {t_* \over t_{\rm max}}$, distinguishing the two time regimes, is identified with the value corresponding to $f_*$ in Fig.~\ref{fig:ftaumu} as $\tilde{t}_* = 0.675\pm 0.075, 0.55\pm 0.07, 0.505\pm 0.025$, and $0.395\pm 0.025$ for $\mu  = 1, 0.1, 0.05$, and $0.02$, respectively. 
\item The degree distributions of the compounds in the bipartite metabolic networks of individual species are in good agreement with the empirical distribution for all $\mu$, which are approximated by a power-law with the exponent close to $2$ [Fig.~\ref{fig:Pkmu}].
\item The crossover behaviors of the first-recruitment time distribution $P_{\rm rec}(\tau)$ are observed for all $\mu$ with the crossover scale $f_*$ increasing with decreasing $\mu$. The probability $\beta(\tau)$ that the reaction recruited at time $\tau$ by a species is brand-new decays weakly with the number of species $S(\tau)$ in the early-time regime and decays inversely with $S(\tau)$ in the late-time regime for all $\mu$. 
\item  The popularity $f_r$ of a reaction $r$ at the simulation time $t_{\rm max}$ decreases with its first-recruitment time $\tau_r$ exponentially for all $\mu$ [Fig.~\ref{fig:ftaumu}]. Its time derivative in the late-time regime is approximately ${df_r\over d\tau_r}\sim - f_r^{0.8}$ with the exponent varying little with $\mu$. 
\end{itemize}

\section{Distribution of the distance between species regarding their sets of reactions}

The phylogenetic tree represents the collection of the parent-daughter relationship of all the species. The comparison of the phylogenetic tree between the empirical data and the simulation of our model cannot be made directly since the identity of each species is not preserved between the two. The phylogenetic tree can be constructed by performing the hierarchical clustering with the distance between every pair, and we are interested in the phylogenetic tree constructed by the distance of the sets of reactions of two species. If the distance matrix is similar between empirical data and simulations, then the phylogenetic tree will be similar in structure. 

We introduce the metabolic distance $d_{s_1, s_2}$ between two species $s_1$ and $s_2$ in terms of the Jaccard index between their sets of metabolic reactions 
\begin{equation}
d_{s_1,s_2} \equiv 1 - J_{s_1,s_2} {\rm with} \ J_{s_1,s_2} \equiv {|\cR^{s_1} \cap \cR^{s_2}|\over |\cR^{s_1} \cup \cR^{s_2}|}={\sum_r A^{s_1}_r A^{s_2}_r \over \sum_r \theta(A^{s_1}_r + A^{s_2}_r)},
\label{eq:d12}
\end{equation}
where $|\cR^s|$ represents the number of elements of the set $\cR^s$ of reactions of species $s$, $A^s_r=0, 1$ is the species-reaction association matrix, and $\theta(x)=1$ if $x>0$ and $\theta(x)=0$ otherwise.
The structure of the obtained phylogenetic tree should depend on the distribution of the metabolic distance between species, 
\begin{equation}
P_{\rm dist}(d) = {2\over S(S-1)} \sum_{i=1}^S \sum_{j=i+1}^S \delta(d_{s_i, s_j}-d).
\end{equation}
We compare $P^{\rm (real)}_{\rm dist}(d)$ and  $P^{\rm (model)}_{\rm dist}(d)$ in Fig. 3(d).

\end{document}